\documentclass[12pt]{article}

\usepackage{amsmath,amsfonts,amssymb}
\usepackage{epsfig}
\usepackage{array}

\newtheorem{Def}{Definition}[section]
\newtheorem{Rm}{Remark}[section]
\newtheorem{Prop}{Proposition}[section]

\def\sectappend#1
{\vskip35pt plus12pt
{\raggedright
\noindent {\Large \bf {#1}\par}}
\vskip13pt}

\def\build#1_#2^#3{\mathrel{\mathop{\kern 0pt#1}\limits_{#2}^{#3}}}

\def\beq{\begin{equation}}
\def\eeq{\end{equation}}
\def\beqn{\begin{eqnarray}}
\def\eeqn{\end{eqnarray}}

\expandafter\ifx\csname amssym.def\endcsname\relax \else\endinput\fi

\expandafter\edef\csname amssym.def\endcsname{
       \catcode`\noexpand\@=\the\catcode`\@\space}

\catcode`\@=11

\def\undefine#1{\let#1\undefined}
\def\newsymbol#1#2#3#4#5{\let\next@\relax
 \ifnum#2=\@ne\let\next@\msafam@\else
 \ifnum#2=\tw@\let\next@\msbfam@\fi\fi
 \mathchardef#1="#3\next@#4#5}
\def\mathhexbox@#1#2#3{\relax
 \ifmmode\mathpalette{}{\m@th\mathchar"#1#2#3}
 \else\leavevmode\hbox{$\m@th\mathchar"#1#2#3$}\fi}
\def\hexnumber@#1{\ifcase#1 0\or 1\or 2\or 3\or 4\or 5\or 6\or 7\or 8\or
 9\or A\or B\or C\or D\or E\or F\fi}

\font\tenmsa=msam10 scaled \magstep1
\font\sevenmsa=msam7 scaled \magstep1
\font\fivemsa=msam5 scaled \magstep1
\newfam\msafam
\textfont\msafam=\tenmsa
\scriptfont\msafam=\sevenmsa
\scriptscriptfont\msafam=\fivemsa
\edef\msafam@{\hexnumber@\msafam}

\mathchardef\dabar@"0\msafam@39
\def\dashrightarrow{\mathrel{\dabar@\dabar@\mathchar"0\msafam@4B}}
\def\dashleftarrow{\mathrel{\mathchar"0\msafam@4C\dabar@\dabar@}}

\def\ulcorner{\delimiter"4\msafam@70\msafam@70 }
\def\urcorner{\delimiter"5\msafam@71\msafam@71 }
\def\llcorner{\delimiter"4\msafam@78\msafam@78 }
\def\lrcorner{\delimiter"5\msafam@79\msafam@79 }
\def\yen{{\mathhexbox@\msafam@55 }}
\def\checkmark{{\mathhexbox@\msafam@58 }}
\def\circledR{{\mathhexbox@\msafam@72 }}
\def\maltese{{\mathhexbox@\msafam@7A }}

\font\tenmsb=msbm10 scaled \magstep1
\font\sevenmsb=msbm7 scaled \magstep1
\font\fivemsb=msbm5 scaled \magstep1
\newfam\msbfam
\textfont\msbfam=\tenmsb
\scriptfont\msbfam=\sevenmsb
\scriptscriptfont\msbfam=\fivemsb
\edef\msbfam@{\hexnumber@\msbfam}
\def\Bbb#1{{\fam\msbfam\relax#1}}

\def\widehat#1{\setbox\z@\hbox{$\m@th#1$}
 \ifdim\wd\z@>\tw@ em\mathaccent"0\msbfam@5B{#1}
 \else\mathaccent"0362{#1}\fi}
\def\widetilde#1{\setbox\z@\hbox{$\m@th#1$}
 \ifdim\wd\z@>\tw@ em\mathaccent"0\msbfam@5D{#1}
 \else\mathaccent"0365{#1}\fi}

\font\teneufm=eufm10 scaled \magstep1
\font\seveneufm=eufm7 scaled \magstep1
\font\fiveeufm=eufm5 scaled \magstep1
\newfam\eufmfam
\textfont\eufmfam=\teneufm
\scriptfont\eufmfam=\seveneufm
\scriptscriptfont\eufmfam=\fiveeufm

\csname amssym.def\endcsname

\def\ben{\begin{enumerate}}
\def\een{\end{enumerate}}
\def\bit{\begin{itemize}}
\def\eit{\end{itemize}}
\def\im{\vspace{-0.25 cm}\item[-]}

\def\b{\beta}

\def\d{\delta}
\def\ep{\epsilon}

\def\s{\sigma}
\def\bS{\boldsymbol{S}}

\def\bs{\boldsymbol{\sigma}}
\def\bl{\boldsymbol{l}}
\def\bn{\boldsymbol{n}}

\def\L{\Lambda}
\def\S{\Sigma}

\let\rel=\relatif

\def\Prob{\mathop{\rm I\!P} }

\def\bsum{\mathop{\pmb{\sum}}}
\def\bprod{\mathop{\pmb{\prod}}}
\def\bint{\mathop{\pmb{\int}}}

\def\RC{\hspace{-0.2cm} \it RC\hspace{0.1cm}\rm}
\def\RCR{\hspace{-0.2cm} \it RCR\hspace{0.1cm}\rm}
\def\RCM{\hspace{-0.2cm} \it RCM\hspace{0.1cm}\rm}
\def\MC{\it MC\,\rm}

\title{\large \bf
Percolation and Magnetization \\
in the Continuous Spin Ising Model}

\author{\normalsize
Piotr Bialas$^{1,2}$,
Philippe Blanchard$^{1}$,
Santo Fortunato$^{1}$, \\
\normalsize Daniel Gandolfo$^{1,3}$,
Helmut Satz$^{1}$
}

\date{} 

\begin{document}

\maketitle
\begin{quote}
\footnotesize
\begin{center}
$^{1}$ Fakult\"{a}t f\"{u}r Physik, 
Universit\"{a}t Bielefeld, D-33615, Bielefeld, Germany.\\
$^{2}$ Institute of Comp. Science, Jagellonian University, PL-30-072 Krakow, Poland.\\
$^{3}$ D\'ept. de Math., Universit\'e de 
Toulon et du Var, , F-83957 La Garde Cedex,\\ 
\& CPT, CNRS, Luminy, case 907, 13288 Marseille Cedex 09, FRANCE.
\end{center}
\end{quote}

\bibliographystyle{unsrt}


\begin{quote}
{\bf Abstract.}
In the strong coupling limit the partition function 
of $SU(2)$ gauge theory 
can be reduced to that of the
continuous spin Ising model with 
nearest neighbour pair-interactions. 
The random cluster representation of the continuous spin Ising model 
in two dimensions is derived through a
Fortuin-Kasteleyn transformation, and the properties of the corresponding cluster distribution
are analyzed. 
It is shown that for this model,
the magnetic transition is equivalent to  
the percolation transition of Fortuin-Kasteleyn clusters, using
local bond weights. 
These results are also illustrated by means of numerical simulations.
\end{quote}
\section{Introduction}
\setcounter{equation}{0} 

It has recently been proposed that the deconfinement transition in SU(2) gauge
theory can be characterized as percolation of Polyakov loop clusters
\cite{forsatz}. This idea
is based on the fact that SU(2) gauge theory and the Ising model belong to the
same
universality class \cite{Svet}, and that the magnetization transition in the
Ising
model can be specified as percolation of clusters defined through local bond
weights \cite{Fortuin}. It thus seems natural that the same holds
for the corresponding Polyakov loop clusters in SU(2) gauge theory, provided
suitable bond weights can be defined. In \cite{forsatz}, it was shown that in a
lattice formulation effectively corresponding to the strong coupling limit,
SU(2) gauge theory indeed leads to the predicted Ising critical exponents. 

In the present paper, we want to consider the classical continuous spin model on
$\rel^2$ introduced by Griffiths \cite{Gr},
an Ising model with spins taking values continuously between $-1$ and $+1$, and
prove that in this case magnetization and percolation transitions coincide. 
We also present a detailed numerical study of the model with simulations on the
2d square lattice.

For SU(2) lattice gauge theory it was shown in
the strong coupling limit \cite{Ka} that the partition function 
can be written in a form which, apart from a factor which depends on the group
measure,
is the partition function of the classical spin model, with spins varying
continuously in some bounded real interval \cite{Gr}. These continuous spins 
thus are the natural counterparts of the Polyakov loops in SU(2) gauge theory.
Following the pioneering work \cite{Fortuin}, the relationship between the
thermodynamical features of 
spin models and the properties of the corresponding geometrical clusters have in
the past years received considerable attention (see \cite{SW} - \cite{Cha2}).
We want to address here the random percolation content of the continuous 
spin Ising model \cite{Gr} obtained through a Fortuin-Kasteleyn
transformation \cite{Fortuin} of the partition function. A specific class
of geometrical clusters \cite{Wo} shows a percolation transition 
whose critical behaviour matches the conventional Ising counterparts. In
particular, it can be shown that up to finite constants, the spin magnetization
equals the probability of long range cluster connectivity; moreover, the spin
susceptibility becomes equal (up to the same constants) to some mean cluster
size \cite{BCG}. 

The numerical method to be used here is the Wolff algorithm \cite{Wo}. 
It is based on the formulation of a single cluster update algorithm for spin
systems with unprecedented performances regarding the problem of
critical slowing down behaviour near phase transition \cite{Wo}. It turns out
that the Wolff clusters exhibit critical percolation behaviour at the point 
of the Ising phase transition. Very recently, this specific property
of Wolff clusters has been put into a rigorous framework \cite{Cha1},
leading to a concept of Wolff measures which provides the theoretical basis for the 
method to be used below. We shall extend this work to 
the 2-d continuous spin model and corroborate the results through numerical
simulations.

\section{The Model.}
\setcounter{equation}{0}

Let $\L \subset \rel^2$ be a finite lattice with edge set $E$ and
vertex set $V=\{1,..,N\}$ and $\mathbf{S}$ a random configuration of spin
variables
$\mathbf{S}=\{S_{i}\}\subset \S^{V}$, where the $S_{i,\, i\in V}$ take
values independently in $\S=[-1,+1]$.
Consider the Hamiltonian 
\begin{equation}
\displaystyle H(\mathbf{S})=-\mathop{\bsum}_{<i,j>} J_{i,j} S_{i} S_{j},
\end{equation}
where the sum is over all nearest neighbour pairs ($<i,j>$) of spins
and $J_{i,j}>0$.
Rewriting the spin variables $S_{i}=l_{i} \s_{i}$, where $\s_{i} = sign(S_{i})
=\pm1$ and $l_{i} = |S_{i}| \in [0,1]$, one has
\begin{equation}
\displaystyle H(\mathbf{S}) = H(\bs,\bl) =
-\mathop{\bsum}_{i,j} J_{i,j} l_{i} l_{j} \s_{i} \s_{j}
\label{ham}
\end{equation}

\subsection{Joint Distribution over Spin Amplitudes and Bond Variables}

Following Edwards and Sokal \cite{ES}, the random cluster representation
(\RCR\hspace{-0.1cm}) \cite{Fortuin,G} of this model is derived in the following
way.
Let $\bn \subset \{0,1\}^{E}$ be a configuration of bond variables on
$\L$ where,
\smallskip
\ben
\im $n_{ij}=1$ means that there is a bond between sites $i$ and $j$ in $V$,
\im $n_{ij}=0$ is the opposite event.
\een
Consider the following joint probability distribution of the random variables
$(\s,l, n)$,
\begin{eqnarray}
\displaystyle\Prob(\mathbf{S}, \bn) = \Prob(\bl,
\bs, \bn) &=& Z_{\Lambda}^{-1} \mathop{\bprod}_{<i,j>}
e^{\beta J_{i,j} l_{i} l_{j}}  \left[p_{i,j} \d_{n_{ij,1} } \d_{\s_{i}
\s_{j} }  +
(1-p_{ij}) \d_{n_{ij,0}} \right] \nonumber \\
&=& Z_{\Lambda}^{-1} \mathop{\bprod}_{<i,j>} w_{\beta,J} (\bl, \bs,\bn),
\label{jointprob}
\end{eqnarray}
where $p_{i,j}=1-e^{- 2 \beta J_{i,j} l_{i} l_{j}}$ is a weighting factor
to be interpreted as the probability for a link $n_{ij}$ to exist ($n_{ij}=1$)
between two nearest neighbour spins $S_{i}$ and $S_{j}$ that share the same
sign ($\d_{\s_{i} \s_{j}}=1$), and
$(1-p_{i,j})$ the probability of having no link ($n_{ij}=0$) between them.
$Z_{\Lambda}$ is the partition function of the model given by
\begin{equation}
\displaystyle Z_{\Lambda}  =
\mathop{\bsum}_{\{\s\}} \bint \mathop{\bprod}_{i} df_{i} (l_{i} )
\exp{ [ \beta \displaystyle \mathop{\bsum}_{<i,j>} J_{i,j} l_{i} l_{j} \s_{i}
\s_{j} ] },
\label{RCM}
\end{equation}
where $df_{i}$ is the probability distribution of the spin amplitudes $l_{i}$ and
$\beta$ is the inverse temperature.

Starting with the joint distribution (\ref{jointprob}), it is
straightforward to express the
weights $\displaystyle w_{\beta,J} (\bl, \bs,\bn)$ in the following way
$$\displaystyle w_{\beta,J}(\bl, \bs,\bn) =
\mathop{\bprod}_{<i,j>} e^{\beta J_{i,j} l_{i} l_{j}}
\mathop{ \mathop{\bprod}_{<i,j>}}_{n_{ij}=1} p_{i,j}  \d_{\s_{i} \s_{j}}
\mathop{ \mathop{\bprod}_{<i,j>}}_{n_{ij}=0} (1-p_{ij}).
$$
Now, we write $\tilde{\bn}$ for a bond configuration fulfilling the
compatibility condition that {\it a bond $\tilde{\bn}_{i,j}$ exists
between sites $i$ and $j$ ($\tilde{\bn}_{i,j}=1$) iff $\s_{i} = \s_{j}$ }
and call
$c(\tilde{\bn})$ the
number of clusters of bonds in the configuration $\tilde{\bn}$.
Then, the weights $w_{\beta,J}$ take the form (FK-representation)
\begin{equation}
\displaystyle w_{\beta,J}(\bl, \tilde{\bn}) =
\mathop{\bprod}_{<i,j>} e^{\beta J_{i,j} l_{i} l_{j}} \left(
2^{c(\tilde{\bn})} \mathop{\bprod}_{<i,j>\in\tilde{\bn}} p_{i,j}
\mathop{\bprod}_{<i,j>\notin\tilde{\bn}} (1-p_{ij}) \right).
\label{Wo}
\end{equation}
These weights are called the Wolff weights of the random cluster distribution.
The joint distribution on the configurations
($\bl, \tilde{\bn}$) provide the right framework to relate
the critical behavior in the original spin system and its associated
percolation representation.

\Rm{The Wolff weights $w_{\beta,J}(\bl, \tilde{\bn})$ differ from the (FK)
 weights of the Ising case (see \cite{Fortuin,CK}) by the term $\displaystyle \mathop{\bprod}_{<i,j>}
 e^{\beta J_{i,j} l_{i} l_{j}}$ that reflect the measure on the spin amplitudes in the continuous case.}
\rm


\section{Wolff Cluster Algorithm and Distributions}
\setcounter{equation}{0}
Non-local cluster Monte Carlo (\MC) algorithms have brought significant
improvements in the simulation of Ising models near criticality. Starting from
the ground-breaking work of Fortuin and Kasteleyn \cite{Fortuin}, which relates the
partition function of spin systems with that of a correlated percolation
model,
Swendsen and Wang \cite{SW} derived a non-local cluster algorithm
which drastically reduces the critical slowing down phenomena near the
transition point, with a dynamical critical exponent near $0.25$, where $z$
is defined by
\begin{equation}
\tau=\xi^{z}.
\end{equation}
Here $\tau$ is the correlation time in \it MC \rm simulations (measured in
\it MC \rm steps per
site) and $\xi$ is the correlation length. For a system of size $L$, near the
critical point, $\xi$ scales like $L^2$ in two dimensions. In the case of
a local update algorithm like Metropolis or Heatbath, $z$ is found to
be close to $2$, so the
required time to reach stable configurations at criticality in this case is of
order $L^4$, reducing noticeably the possible size of samples to study.

Building on the Swendsen-Wang idea, Wolff \cite{Wo} improved the method (see below)
and derived a non-local update \it MC \rm  scheme with a dynamical critical
exponent $z$ smaller than or equal to Swendsen-Wang's in any dimension and
furthermore easier to implement.

\subsection{The Wolff Algorithm}
First let us recall briefly the main features of the Wolff method.
Consider the $2$-$D$ ferromagnetic Ising Model on $\rel^2$ with coupling
constant $J$ and inverse
temperature $\b$.
Starting from a randomly chosen spin $\s_0$, visit all nearest neighbours and
\medskip
\ben
\im with probability $p_{b}=1-exp(-2\b J)$,
\im and only if they have the same orientation as $\s_0$, \hfill \mbox{(III.1)}
\label{CC}
\een
include them in the same cluster as $\s_0$; spins not satisfying both
conditions are excluded. Repeat
iteratively this
procedure with newly added spins in the cluster
until no more neighbours fulfill the above compatibility condition
($III.1$). Now flip all the spins
in that cluster with {\it probability} $1$. After that, erase all the bonds
and start
this procedure again. It turns out that this
dynamics verifies the detailed balance condition, i.e. it samples the Gibbs
distribution of the Ising model (see \cite{Wo}). The distinguishing feature
of the Wolff
method compared to Swendsen-Wang's is that, in the latter (following
Fortuin and Kasteleyn \cite{Fortuin}) one needs to build, with the same growing
probability $p_{b}$
as before, {\it all possible clusters} of like spins and then, with
probability 1/2, flip all the spins in those clusters. Then all the
bonds are erased and one starts again from the newly created spin
configuration.
It can also be shown that this method verifies the detailed balance
condition \cite{SW}.
In summary, the following remarks can be made:
\medskip
\ben
\im The building procedure of an individual cluster in both methods is clearly
    the same, thus the Wolff cluster belongs to the set of Swendsen-Wang
clusters.
\im When the Wolff cluster is built, the randomly chosen spin has obviously
    higher probability to fall in a large Swendsen-Wang cluster than in a
smaller one.
\een
It results that the distribution of Wolff clusters is given by the
distribution of
Swendsen-Wang clusters (see below) modified by an additional weight that
takes into
account the size of the clusters.
\Rm{A (rather non economic) way of building the Wolff cluster would be to
construct the set of all $\cal N$ Swendsen-Wang bond clusters ${\cal
C}^{SW}_{i}$,
$i\in \{1,..,{\cal N}\}$, then to pick at random a lattice spin $\s_{\bf x}$
and flip with probability $1$ all the spins belonging to the cluster ${\cal
C}^{SW}(\s_{\bf x})$.
Obviously, the randomly chosen spin $\s_{\bf x}$ has higher probability to sit
in a large cluster}. \rm

\Rm{In the case of our model, the bond probability
  $p_{b}(S_{i},S_{j})=1-exp(-2 \b J_{i,j}l_{i}l_{j})$  takes also into
account the
  spin amplitudes. It turns out that, the Wolff dynamics (as
  described above) is no longer ergodic (because spin amplitudes are
locally conserved)
  and must be supplemented by a local update method (like Metropolis or
Heatbath)
  in order to respect detailed balance. In practice it is sufficient to
insert a Heatbath sweep
  after several Wolff sweeps have been performed. This of course slightly
alters
  the gain brought by the non-local cluster update method concerning
  critical slowing down near the phase transition.} \rm

\medskip
The proof that the Wolff algorithm applied to our infinite-spin model fulfills
both ergodicity criteria and detailed balance conditions will be omitted here
since it follows closely the derivations that can be found in
\cite{Wo2,Ha,Cha3,Cha4}.

\subsection{Conditional Distributions and Properties of Wolff Clusters}
We state in this section our main results concerning the properties of the
\RC distribution.

First, it is straightforward to see that integrating $\Prob(\bl, \bs, \bn)$
over $\bn$
  gives the marginal distribution of the model with Hamiltonian
  (\ref{ham}).

\Prop{The marginal spin distribution in (\ref{jointprob}) gives the Boltzmann
  weight associated with Hamiltonian (\ref{ham}).}
\rm

Tracing over bond variables in (\ref{jointprob}), one gets
\begin{eqnarray}
\mathop{\bsum}_{{n_{ij}=0}}^{1} \Prob(\bl,\bs,\bn) & = &
Z_{\Lambda}^{-1} \mathop{\bprod}_{<i,j>} \mathop{\bsum}_{{n_{ij}=0}}^{1}
\left\{ e^{\beta J_{i,j}  l_{i} l_{j}}  \left[ p_{i,j} \d_{n_{ij,1} }
\d_{\s_{i} \s_{j} }  +
(1-p_{ij} )\d_{n_{ij,0}} \right] \right\} \nonumber \\
& = & Z_{\Lambda}^{-1} \mathop{\bprod}_{<i,j>}
\left\{ e^{\beta J_{i,j}  l_{i} l_{j}}  \left[ p_{i,j}  \d_{\s_{i} \s_{j} }
+ (1-p_{ij}) \right] \right\}
\nonumber
\end{eqnarray}
using the expression for $p_{i,j}$ and re-arranging terms, this can be
rewritten as
\begin{eqnarray}
& Z_{\Lambda}^{-1} & \mathop{\bprod}_{<i,j>}
\left\{ e^{\beta J_{i,j}  l_{i} l_{j}} \d_{\s_{i} \s_{j}} + e^{- \beta
J_{i,j}  l_{i} l_{j}} (1-\d_{\s_{i}
\s_{j}}) \right\} \nonumber \\
= & Z_{\Lambda}^{-1} & \mathop{\bprod}_{<i,j>}
\left\{ e^{\beta J_{i,j}  l_{i} l_{j} \s_{i} \s_{j}} \d_{\s_{i} \s_{j}} +
e^{\beta J_{i,j}  l_{i} l_{j} \s_{i} \s_{j}} (1-\d_{\s_{i}
\s_{j}}) \right\} \nonumber
\end{eqnarray}
thus we get for this marginal distribution
\begin{equation}
Z_{\Lambda}^{-1} \mathop{\bprod}_{<i,j>}
e^{\beta J_{i,j}  l_{i} l_{j} \s_{i} \s_{j}} =
Z_{\Lambda}^{-1} \exp{\big(\mathop{\bsum}_{<i,j>} \beta J_{i,j} l_{i} l_{j}
\s_{i} \s_{j}  \big)}  =
Z_{\Lambda}^{-1} \exp{\big(\mathop{\bsum}_{<i,j>} \beta J_{i,j} S_{i} S_{j}
\big)}  =
\frac{e^{H(\bS)}}{Z_{\Lambda}},
\nonumber
\end{equation}
which is just the Boltzmann weight for the spin configuration $\bS$ of our
model.

\smallskip
The nice features that make Wolff clusters useful reside in the following
propositions:

\Prop{The magnetization $M$ of the spin system is equal to the probability of
  long range connectivity in the percolation model.
\label{M=P}}
\rm

Considering the joint bond-spin distribution (\ref{jointprob}) with weights
(\ref{Wo}), one can ask: given a bond configuration, what is the conditional
distribution on the spins ?
Let us assume ($+$) boundary conditions for the spins (which amounts at fixing
$l_i \s_i = 1$ on the boundary).
First, due to the compatibility condition ($III.1$), if site $k$ is
connected to the
boundary through some path in the percolation model, then necessarily
$\s_k=+1$. The same is
true with ($-$) boundary conditions due to the sign symmetry of the
exponential factor
in the statistical weight. Second, a site that is not connected to the boundary
is equally likely to be in a $\pm 1$ state. From this it follows that $M$ is
exactly given by the
probability of connectivity to the boundary, thence long range order in the
spin system is the same as percolation in the random cluster model.

A more detailed argument will be found in \cite{BCG}. It is actually easy to
show that absence of percolation is equivalent to zero magnetization.
Consider again the random cluster model (\RCM) characterized by the
joint distribution (\ref{jointprob}). First let us state the

\Def{Let $\Prob_{\L,i}(\b)$ be the probability for a site $i$ to be
connected to the
  boundary of a finite domain $\L\subset \rel^{2}$ through some path of active
bonds in the percolation model. We say that there is no percolation in the
{\RCM} iff
\begin{equation}
\forall i\in\L, \; \lim_{\L\to\rel^{2}} {\Prob}_{\L,i}(\b) =
{\Prob}_{\infty,i}(\b) = 0
\nonumber
\end{equation}
\label{inflim}
}\rm
\vspace{-0.5cm}

\Prop{Consider the joint bond-spin distribution (\ref{jointprob}) and
suppose that for the infinite-spin model at inverse temperature $\b$ there
corresponds no percolation in the joint correlated geometrical bond
representation; then the
 magnetization vanishes.
}\rm

The argument is as follows. Let $\bs_{n}=(\s_{i_1},\s_{i_2},..,\s_{i_n})$ a
finite set of spins and
$M_{\L}^{(\mathfrak{B})}(\bs_{n})$ the corresponding magnetization in the
spin model with boundary conditions $(\mathfrak{B})$.
Let $(\mathfrak{B}-\ep)$, the boundary condition identical to
$(\mathfrak{B})$ except that
each spin value on the boundary has been lowered by an amount $\ep$. Then one has
\begin{equation}
\langle M_{\L}^{(\mathfrak{B})}(\bs_{n}+\ep) \rangle \equiv \langle
M_{\L}^{(\mathfrak{B}-\ep)}(\bs_{n}) \rangle
\nonumber
\end{equation}
where $\bs_{n}+\ep=(\s_{i_1}+\ep,\s_{i_2}+\ep,..,\s_{i_n}+\ep)$.
Clearly the geometrical clusters corresponding to these two boundary conditions
are the same, thus for all bond configurations in which sites
${i_1},{i_2},..,{i_n}$ are not connected to the boundary, the corresponding
contributions to the magnetization are the same.

Now, using the uniqueness of the limiting state at given $\b$, it is easy to
conclude that, in the infinite volume limit of
definition (\ref{inflim}), absence of percolation is the same as zero
magnetization.

From the proposition above follow also interesting results concerning high
temperature decay of correlations for functions of the spin variables
when their relative distance goes to infinity but we will not enter this
question here
and refer to \cite{BCG}.

The stronger statement of proposition (\ref{M=P}) whose
details of proof will also be found in the last reference (see also
\cite{Cha1,Cha2}) is the following:
not only the onset of percolation is the signature of the magnetic ordering but
moreover, there exist a constant $C$ depending only smoothly on $\b$ such that 
\begin{equation}
{\Prob}_{\infty}(\beta) \; \ge \; M(\beta) \; \ge \; C \; {\Prob}_{\infty}(\beta)
\label{m=p}
\end{equation}
Explicitly, there is spontaneous magnetization if and only if there is
percolation and the critical behaviour is the same for these two quantities.
Namely, considering on one hand the order parameter exponent $\b$, defined for the spin system by $M\sim
(T_{c}-T)^{\b}$, where $T_{c}$
is the critical temperature, and on the other hand the order parameter for the
percolation model, i.e. the fraction of sites $\Prob_{\infty}$ in the
percolating
cluster, which behaves like $(p-p_{c})^{\tilde{\b}}$, where $p_{c}$
is the percolation threshold. It follows from (\ref{m=p}) that
$\b=\tilde{\b}$.

A similar result as (\ref{M=P}) can be proven (see \cite{BCG}) between 
the magnetic susceptibility of the continuous spin system and its geometric
counterpart.

In the following we shall carry out
lattice simulations to determine the critical exponents for cluster
percolation, to illustrate that these indeed lead to the values of
the Ising universality class.

\section{Numerical results}
\setcounter{equation}{0} 

We have performed extensive simulations of our model, choosing six different
lattice sizes, namely $64^2$, $96^2$, $128^2$,
$160^2$, $240^2$ and $300^2$.

Our update step consisted of one heatbath for the 
spin amplitudes and three 
Wolff flippings for the signs, which turned out 
to be a good compromise to reduce sensibly the correlation
of the data without making the move be too much time-consuming.
Every five updates we measured the quantities of interest.
The physical quantities are the energy and the magnetization of 
the system. The percolation quantities have been evaluated
in this way: for a given configuration we consider
the sign of the spins corresponding to the sign of the 
global magnetization, say up. Then we form clusters of   
spins up using the local bond weights $1-exp(-2(J/kT){s_i}{s_j})$.
Once we embody all spins up in clusters we assign to each 
cluster a size $s$ which is the number of spins belonging 
to it. 
We say that a cluster percolates if it spans the lattice in both 
directions, that is if it touches all four sides of the lattice.
This choice was made
to avoid the possibility that, due to the finite lattice size,
one could find more than one percolating cluster, making 
ambiguous the evaluation of our variables.   

Finally we assign to the percolation probability P the value of the
cluster size of the percolating cluster divided by the
number of lattice sites ($P=0$ if there is no
percolation) and to the average cluster size S the value 
of the expression

\begin{eqnarray*}
  S={\sum_{s} \Bigg( \frac{n_{s}s^2}{\sum_{s}{n_{s}s}} \Bigg)}~,
\end{eqnarray*}
where $n_{s}$ is the number of clusters of size $s$
and the sums exclude the percolating cluster.

After some preliminary scans of our program for several 
values of the temperature $\beta$ ($\beta=J/kT$), we focused on
the beta range between $1.07$ and $1.11$, where 
the transition seems to take place. 
The number of iterations for each run goes from 
20000 (for $\beta$ values close to the extremes of the range) to 50000  
(around the center of the range).
We interpolated our data using the density of states method (DSM)
\cite{DSM}.

To locate the critical point of the physical transition 
we used the Binder cumulants 
\begin{eqnarray*}
  g_r=\frac{\langle{M^4}\rangle}{{\langle{M^2}\rangle}^2}-3~.
\end{eqnarray*}
Figure 1 shows $g_r$ as a function of $\beta$ for the different
lattice sizes we used. The lines cross remarkably 
at the same point, which suggests that also in our case $g_r$ 
is a scaling function. As a numerical proof we replot our lines
as a function of $(\beta-\beta_{crit})L^{1/\nu}$, choosing
for the exponent $\nu$ the Onsager value 1. The plot (Figure 2)
shows that indeed $g_r$ is a scaling function. 

\begin{figure}[p]
\begin{center}\epsfig{file=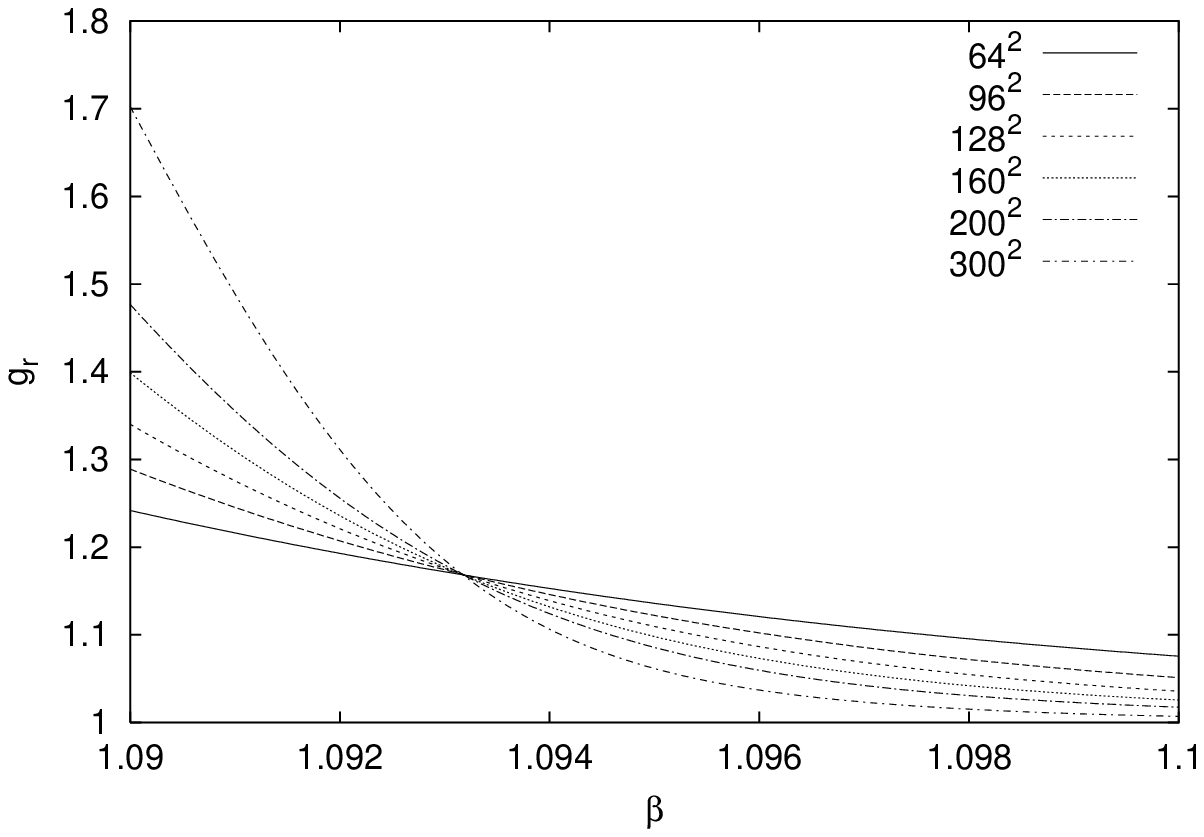,width=13cm}\end{center}
\vspace{-.8cm}
\caption{Binder cumulants as a function of $\beta$ for our six lattice sizes.}
\vspace{1.5cm}
\begin{center}\epsfig{file=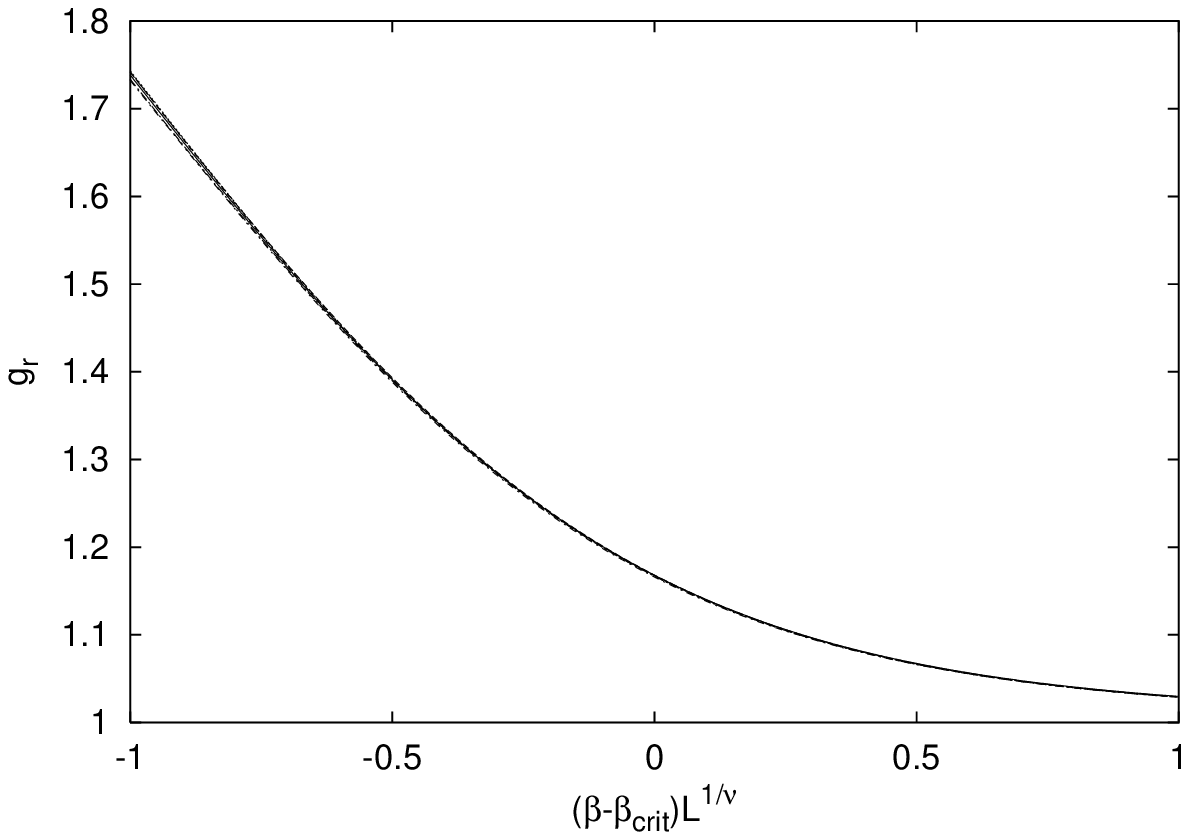,width=13cm}\end{center}
\vspace{-.8cm}
\caption{Rescaled Binder cumulants. We took ${\beta_{crit}}=1.0932$
and for the exponent $\nu$ the Onsager value 1. }
\end{figure}

To get the critical point of the percolation transition we used 
the method suggested in \cite{Bin}, which is based on
the same principle. For each $\beta$ value 
we count the number of configurations where we 
found a percolating cluster and we divide it by
the total number of configurations. If we plot the results as 
a function of $\beta$ for the different lattice sizes 
the corresponding lines should cross at the threshold of the 
geometrical transition (Fig. 3).

\begin{figure}[h]
\begin{center}\epsfig{file=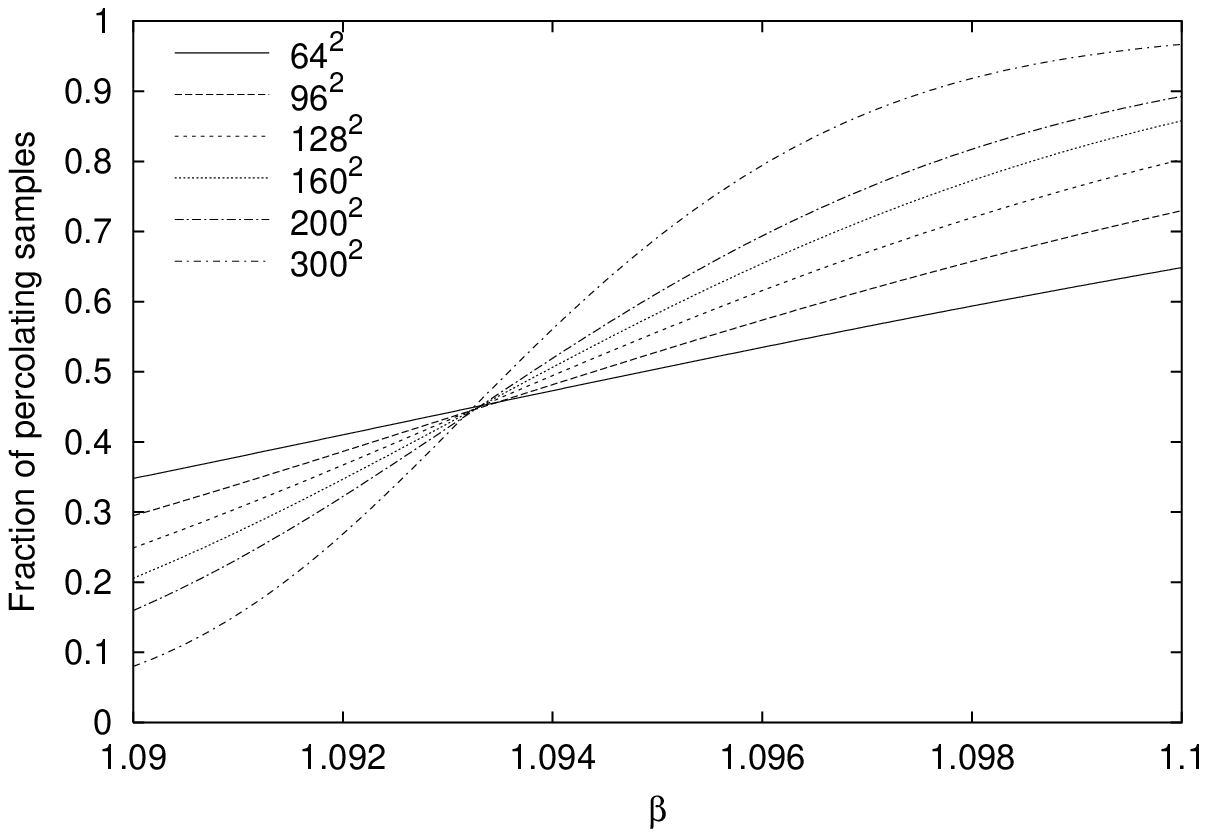,width=13cm}\end{center}
\vspace{-.8cm}
\caption{Fraction of percolating samples as a function of $\beta$ 
for our six lattice sizes.}
\end{figure}

As the figures show, the agreement between the two thresholds we found is excellent.

For the evaluation of the exponents we used the $\chi^2$ method \cite{EMSZ}.
The results we got are reported in Table I. 

  \begin{center}{
      \begin{tabular}{|r|c|c|c|}
\multicolumn{4}{c}{Table I}\\
\hline
  &      Critical point  & $\beta/{\nu}$ & ${\gamma}/{\nu}$ \\ \hline\hline
$\vphantom{\displaystyle\frac{1}{1}}$  Thermal results &$1.093120^{+0.000120}_{-0.000080}$&$
  0.128^{+0.005}_{-0.006}$&$1.745^{+0.007}_{-0.007}$\\  
\hline $\vphantom{\displaystyle\frac{1}{1}}$ 
  Percol. results &$1.093200^{+0.000080}_{-0.000080}$& $ 0.130^{+0.028}_{-0.029}$ &$ 1.753^{+0.006}_{-0.006}$ \\ \hline
      \end{tabular}
      }
  \end{center}

The two thermal exponents' ratios are in good agreement with the
Onsager values.
Unfortunately we 
encountered some troubles in deriving the percolation ${\beta}/{\nu}$ exponents' 
ratio because the best fit values change rapidly in the little range 
of beta values for which the $\chi^2$ corresponds to
the 95\% confidence level. Consequently, the relative error of
${\beta}/{\nu}$ turns out to be large. For a more precise evaluation
of this ratio much higher statistics or higher lattice sizes 
seem to be necessary.

However, the value of the other exponents' ratio ${\gamma}/{\nu}$ 
is in good agreement with the corresponding
Onsager value $7/4$ and its error seems to exclude 
the possibility for it to be instead the random percolation
ratio $43/24$.

\section{Conclusions}
\setcounter{equation}{0}
For the continuous spin Ising model, the
percolation transition of Fortuin-Kasteleyn clusters is indeed equivalent to
the magnetic transition due to spontaneous $Z_2$ symmetry breaking.
We have also determined by means of lattice simulations the critical
exponents of the percolation variables and shown that they
belong to the Ising universality class. The simulation
method used here can also be employed in the study of percolation in
$SU(N)$ gauge theories, where analytic proofs do not exist. We expect
that they will eventually allow a general demonstration that the
deconfinement transition in such theories can indeed be defined as
Polyakov cluster percolation.

\section{Acknowledgments}
\setcounter{equation}{0}
D.G. gratefully acknowledges financial support and kind hospitality 
from the BiBoS research center, Department of Physics, University of Bielefeld, 
Germany.
S.F. was supported by the EU-Network ERBFMRX-CT97-0122 and the DFG
Forschergruppe Ka 1198/4-1; he also acknowledges the Centre de Physique
Th\'eorique, C.N.R.S, Luminy, Marseille, France where part of this work was developed.
P.B. was partially supported by Alexander von Humboltd Foundation and KBN grant 2P03B 149 17.
It is our pleasure to thank L. Chayes for discussions and, for two of us,
Ph. B. and D. G., for the joy of collaboration.

\end{document}